# Demonstration of high-power photonic-crystal surface-emitting lasers with 1-kHz-class intrinsic linewidths

Ryohei Morita[1,*], Takuya Inoue[2], Masahiro Yoshida[1], Kentaro Enoki[3], Menaka De Zoysa[2], Kenji Ishizaki[2], and Susumu Noda[1,2]

[1] *Department of Electronic Science and Engineering, Kyoto University, Kyoto-Daigaku-Katsura, Nishikyo-ku, Kyoto 615–8510, Japan*

[2] *Photonics and Electronics Science and Engineering Center, Kyoto University, Kyoto-Daigaku-Katsura, Nishikyo-ku, Kyoto 615–8510, Japan*

3 *High Frequency & Optical Device Works, Mitsubishi Electric Corporation, 4-1 Mizuhara, Itami, Hyogo 664-8641, Japan*

[*]moritar@qoe.kuee.kyoto-u.ac.jp

**Abstract**

Photonic-crystal surface-emitting lasers (PCSELs) are capable of single-mode, high-power lasing over a large resonator area owing to two-dimensional resonance at a singularity point of the photonic band structure. Since the number of photons in the lasing mode in PCSELs are much larger than those in conventional semiconductor lasers, PCSELs are in principle suitable for coherent operation with a narrow spectral linewidth. In this paper, we numerically and experimentally investigate intrinsic spectral linewidths of 1-mm-diameter PCSELs under continuous-wave (CW) operation, and we demonstrate CW operation with 1-kHz-class intrinsic linewidths and 5-W-class output power.

1. **Introduction**

Semiconductor lasers with narrow spectral linewidths and high output powers are in high demand as compact light sources in applications such as long-distance coherent optical communications [1,2] and highly sensitive optical sensing [3,4]. Because intrinsic spectral linewidths of semiconductor lasers generally depend on the coupling of random, spontaneously emitted light into the coherent lasing mode of the laser cavity, the spectral linewidth becomes narrower as the number of photons in the lasing mode increases and the rate of spontaneous emission decreases [5]. To increase the number of photons inside the laser cavity, it is important to increase the volume of the laser cavity (and the injection current accordingly) and to reduce the optical losses in the laser cavity. However, in conventional semiconductor lasers such as distributed feedback (DFB) lasers, the minimum achievable spectral linewidth using these methods is limited to several kHz due to multimode lasing caused by the reduction of the threshold gain difference [6,7]. For even narrower (sub-kHz) linewidths, semiconductor lasers have relied on external optical feedback systems [8,9], but these systems require precise optical alignment of optical feedback optics. It is possible to circumvent such alignment requirements by considering monolithic integration of the semiconductor lasers and optical feedback systems [10-12], but, in this case, simultaneously realizing narrow spectral linewidths and high (watt-class) output powers is difficult.

Photonic-crystal surface-emitting lasers (PCSELs) show promise toward overcoming the above difficulties. PCSELs are surface-emitting semiconductor lasers that can realize both high output powers and narrow beam divergence angles by utilizing two-dimensional (2D) resonance at a singularity point (such as the Γ point) of 2D photonic crystals [13-16]. So far, a general design guideline for realizing single-mode operation with 10-W to 1-kW-class output powers has been established [16, 17], and, following this guideline, 7W- to 50-W-class single-mode operation has been experimentally achieved in 800-μm to 3-mm-diameter PCSELs under continuous wave (CW) condition [16, 18]. Owing to the capability of such large-area single-mode lasing operation, the number of photons inside the laser cavity of PCSELs can be in principle much higher than those in conventional semiconductor lasers, and thus PCSELs are potentially suitable for narrow-linewidth operation. In our previous study, we reported about initial investigations on the spectral linewidth of PCSELs, and we experimentally demonstrated an intrinsic spectral linewidth below 70 kHz in a PCSEL with a lasing diameter of 250 μm and an output power of several tens of mW [19].

In this paper, we theoretically and experimentally investigate intrinsic spectral linewidths of 1-mm-diameter PCSELs with higher output powers. First, we investigate the theoretical linewidths of 1-mm-diameter PCSELs by numerical simulations considering carrier-photon interactions as well as a nonuniform temperature distribution inside the device. Then, we experimentally evaluate the frequency noise spectra of the fabricated 1-mm-diameter PCSELs under continuous-wave operation, and we demonstrate 1-kHz-class intrinsic spectral linewidths with a 5-W-class CW output power.

## 2. Device structure and numerical simulation

Figure 1(a) shows a schematic cross-section of a PCSEL. The active and photonic crystal layers are sandwiched by cladding layers of different conductivity types. The light generated in the active layer is confined vertically by the refractive-index contrast between the active layer and the cladding layers and resonates in the remaining two dimensions in a mode of the photonic crystal layer. The resonant light is then diffracted upward and downward by the photonic crystal layer, where the downward-diffracted light is reflected upward by the distributed Bragg reflector (DBR) on the backside of the device. The upward-diffracted and DBR-reflected light constitute the laser beam emitted from the surface of the PCSEL. The p-side electrode, which determines the size of the lasing area, is 1 mm in diameter. To achieve stable single-mode oscillation throughout this lasing area, double-lattice photonic crystals [16,17] are employed as shown in Fig. 1 (b). This structure consists of two square lattices superposed in the *x*- and *y*-directions with a shift of approximately one quarter of the wavelength in the material. By finely adjusting the hole distance and the size balance between the two holes, the strengths of diffraction at angles of 180° and 90° in the plane of the photonic crystal (that is, the

Hermitian coupling coefficient $\kappa_{1D} + \kappa_{2D-}$ [17]) can be appropriately weakened, which allows the light to spread over a larger area and increases the threshold gain difference between the fundamental mode and the high-order modes. In addition, the radiation constant of the fundamental mode $\alpha_v$ can be controlled by (1) adjusting the hole-size asymmetry of the double-lattice photonic crystal and (2) adjusting the reflection phase of the DBR, which determines the magnitude of the non-Hermitian coupling coefficient $\mu$ [17,19]. As described in the introduction, intrinsic spectral linewidths become narrower as the photon number inside the cavity increases, which here can be realized by reducing $\alpha_v$ of the lasing mode. However, when $\alpha_v$ becomes small relative to other losses such as the in-plane optical loss $\alpha_{//}$ and the material loss $\alpha_0$, the output power decreases proportionally. Therefore, an appropriate value of $\alpha_v$ should be set to simultaneously realize a high output power and a narrow spectral linewidth. In this study, we set the Hermitian and non-Hermitian coupling coefficients of the double-lattice photonic crystal as $\kappa_{1D} + \kappa_{2D-}$ = - 90.4 - 7.72i cm$^{-1}$ and $\mu$ = 86.5 cm$^{-1}$, respectively, so that a moderately low radiation constant ($\alpha_v$ = 9.8 cm$^{-1}$) and much lower in-plane loss ($\alpha_{//}$ = 0.90 cm$^{-1}$) can be obtained.

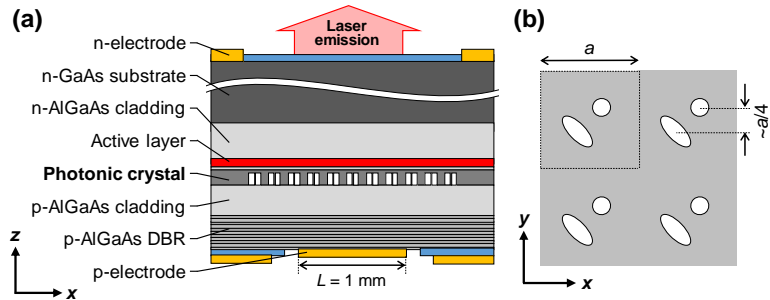

**Fig. 1**. Schematic of a PCSEL. (a) Cross-section of a PCSEL. (b) Schematic of the top surface of a 2D photonic crystal.

To investigate the expected spectral linewidths of a PCSEL of the above design, we perform numerical calculations of the lasing characteristics of a 1-mm-diameter PCSEL using time-dependent three-dimensional coupled wave analysis considering carrier-photon interactions and a nonuniform temperature distribution [19-21]. Figure 2(a) shows the calculated current-light-output characteristics under CW operation. The threshold current is $I_{th}$ ~ 2.9 A, and the slope efficiency is $\eta_s$ ~ 0.89 W/A. A high output power of ~ 5 W is expected with an injection current of ~ 8.0 A. The calculated spectrum at an injection current of 8.0 A is shown in Fig. 2(b). The red line shows a Lorentzian fitting curve with a linewidth of ~ 0.6 kHz, which indicates that the intrinsic spectral linewidth of the PCSEL is indeed sub-kHz. Figure 2(c) shows the calculated intrinsic spectral linewidth as a function of the injection current. As the injection current increases, the intrinsic linewidth becomes narrower, owing to the increase in the number of photons inside the cavity. However, at an injection current of 9.0 A, the calculated linewidth begins to broaden. This broadening is due to an increase in the in-plane loss $\alpha_{//}$ caused by the temperature-induced band-edge frequency change and the resultant mode-gap effect (the details of which are provided in Ref. [19]). From these calculated results, the simultaneous realization of 5-W-class output power and 1-kHz-class intrinsic linewidth is expected of the 1-mm-diameter PCSEL.

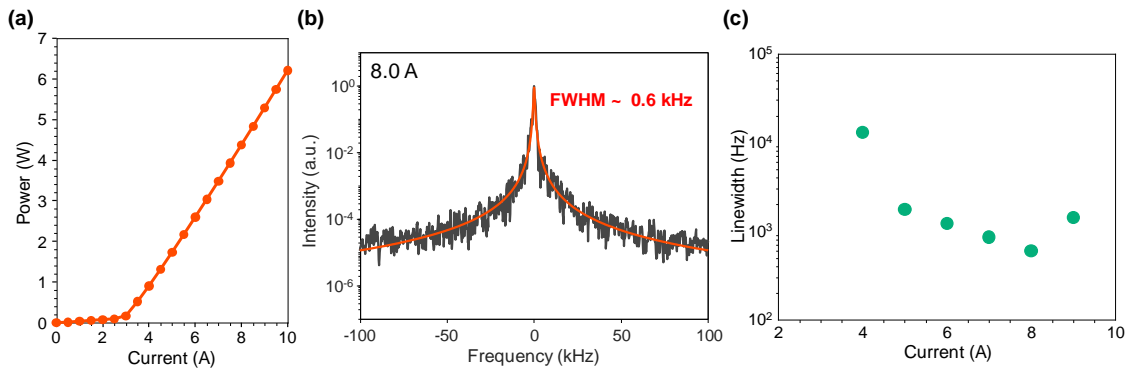

**Fig. 2**. Calculated lasing characteristics of a 1-mm-diameter PCSEL under CW operation. (a) Current-light-output characteristics. (b) Power spectrum at an injection current of 8.0 A. (c) Intrinsic spectral linewidth as a function of the injection current.

3. **Experimental results**

We fabricated a 1-mm diameter double-lattice PCSEL with the same parameters used in the above calculations. Figure 3(a) shows a photograph of the PCSEL mounted on a cooling package. A top-view scanning-electron-microscope (SEM) image of the 2D photonic crystal inside the fabricated PCSEL is shown in Fig. 3(b). The lattice constant was set to $a \sim 276$ nm, and the geometry of each hole was designed to obtain the coupling coefficients used in the numerical calculations above. Figure 3(c) shows the current-light-output characteristics under CW condition. An optical output power of over 5 W was obtained at a current injection of 10.0 A, with a threshold current of $I_{th} \sim 2.9$ A and a slope efficiency of $\eta_s \sim 0.82$ W/A, similar to the numerical results shown in Fig. 2(a). The threshold current density was $J_{th} \sim 0.36$ A/cm$^2$, which is much smaller than that of the 250-μm-diameter PCSEL ($J_{th} \sim 0.63$ A/cm$^2$) used for the linewidth measurements in our previous study [17]. The reasons for the reduction of the threshold current density were that (1) the radiation constant $\alpha_v$ was lower than that of the previous device ($\alpha_v > 20$ cm$^{-1}$), and (2) the in-plane loss of the lasing mode $\alpha_{//}$ was also smaller owing to the larger lasing diameter. The reduction of these losses significantly contributed to an increase in the number of photons inside the laser cavity, resulting in anarrower intrinsic spectral linewidths as shown later. The far-field beam pattern measured at an injection current of 8.0 A is shown in Fig. 3(d). A unimodal beam pattern with a narrow beam divergence angle of $\theta \sim 0.10°$ (evaluated at 1/e$^2$ of the maximum value) was observed. In addition, the lasing spectrum was measured at an injection current of 8.0 A using an optical spectrum analyzer and a single-mode fiber. From this measured lasing spectrum, shown in Fig. 3(e), single-mode oscillation with a lasing wavelength of ~ 940 nm and a high side-mode suppression ratio (SMSR) of > 70 dB was confirmed.

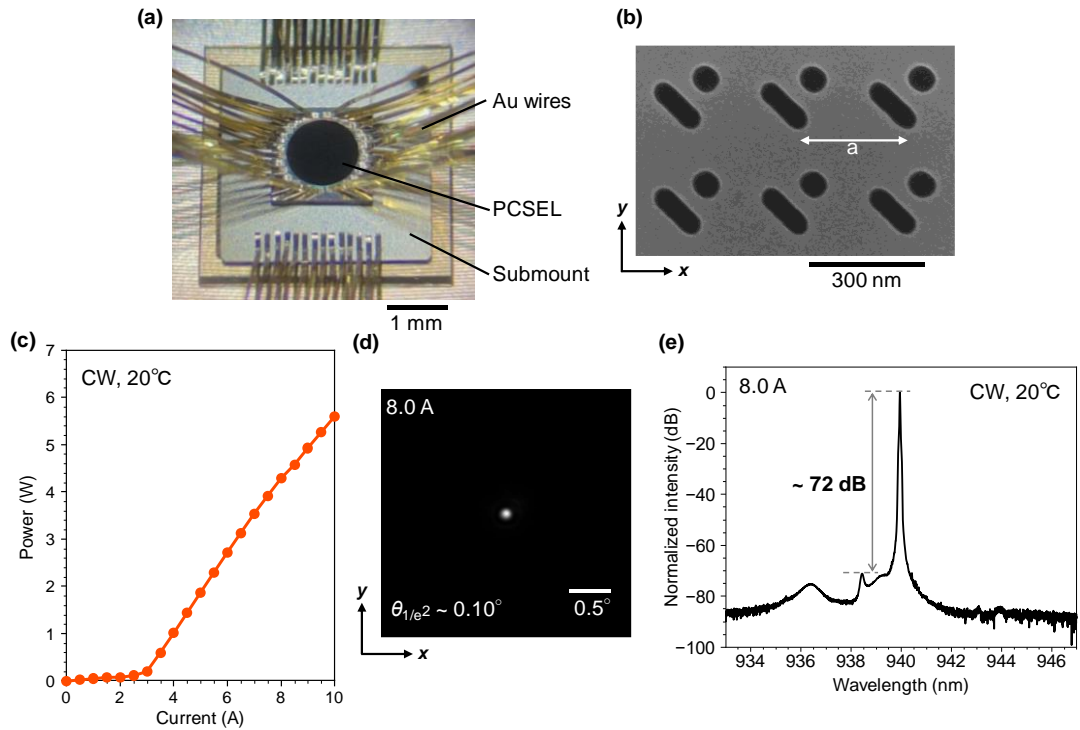

**Fig. 3**. Lasing characteristics of the fabricated PCSEL under continuous-wave operation. (a) Photograph of the fabricated PCSEL with a 1-mm lasing diameter. (b) Top-view scanning electron microscope (SEM) image of the double-lattice photonic crystal inside the fabricated PCSEL. (c) Current-light-output characteristics. (d) Far-field beam pattern at an injection current of 8.0 A. (e) Lasing spectrum at an injection current of 8.0 A.

Figure 4(a) shows a schematic of the experimental setup used for characterization of the intrinsic spectral linewidth of the fabricated PCSEL. The laser beam emitted from the PCSEL, driven by a current source (Thorlabs, ITC4020), was appropriately attenuated by a (90% reflective) beam splitter and a reflective variable attenuation filter, and then coupled into a linewidth analyzer (HighFinesse, LWA-1k-780) via a polarization-maintaining (PM) single-mode fiber. The linewidth analyzer consisted of a frequency discriminator based on an interferometer that converted optical frequency noise into optical intensity noise, and a photodetector that converted optical intensity noise into a voltage signal. The converted voltage signal was Fourier-transformed to obtain a frequency noise power spectrum [22]. The frequency noise power spectrum measured at an injection current of 8.0 A is shown in Fig. 4(b). In the low frequency range, the noise intensity was observed to increase as the frequency decreased, which was mainly due to temporal fluctuations of the injection current in the current source and of the temperature of the PCSEL. In addition, several noise peaks were observed at around 100 kHz, which were considered to be generated from the internal circuit of the current source. On the other hand, in the frequency range above 1 MHz, the noise intensity was almost constant, indicating that white noise due to spontaneous emission was dominant in this range. The intrinsic spectral linewidth can be calculated from the frequency noise power spectral density (PSD) $S_0$ in this flat frequency range as $\Delta v_{int} = \pi S_0$ [23]. The dependence of the intrinsic spectral linewidth on the injection current is shown in Fig. 4(c). As shown in this figure, a 1-kHz-class intrinsic spectral linewidth ($\Delta v_{int} < 1.23$ kHz), corresponding to the lower measurement limit of the linewidth analyzer (~1 kHz), was confirmed at an injection current of 8.0 A. This measured linewidth agrees well with the numerical results shown in Fig. 2(c). More importantly, in addition to this narrow linewidth, our PCSEL simultaneously achieved a high output power of 5 W and a narrow beam divergence angle of 0.1°. A light source capable of such high-power, narrow-linewidth operation will be useful in various applications requiring narrow spectral linewidths and high beam brightnesses, such as long-distance free-space optical communications in space [24] and spaceborne light detection and ranging [25].

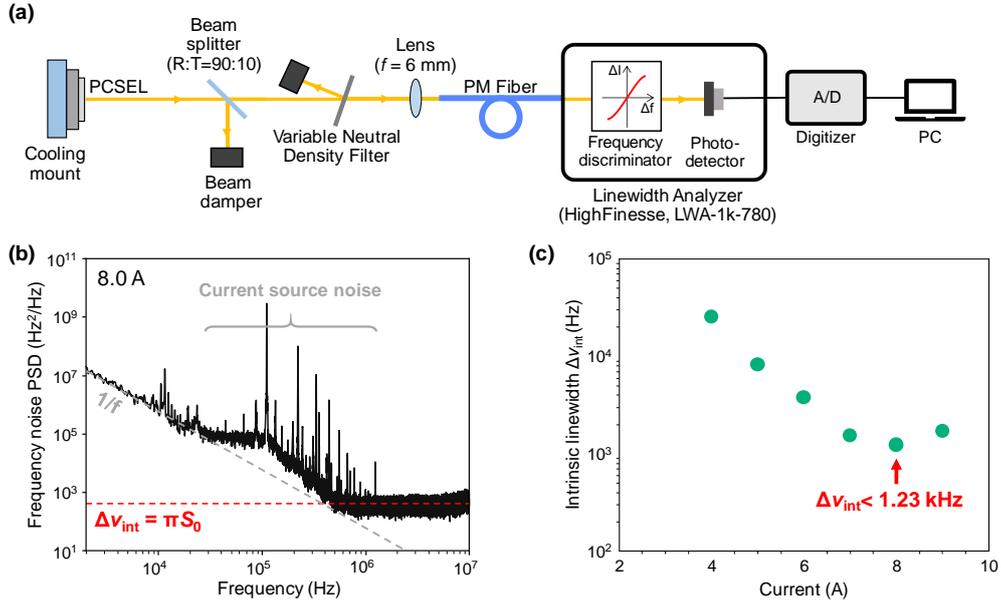

**Fig. 4**. Measurement of the intrinsic spectral linewidths of the fabricated PCSEL. (a) Schematic of the experimental setup for the characterization of the intrinsic spectral linewidths of the fabricated PCSEL. (b) Frequency noise power spectrum at an injection current of 8.0 A. (c) Intrinsic spectral linewidths as a function of the injection current.

## 4. Conclusion

We have theoretically and experimentally investigated the intrinsic spectral linewidth of a PCSEL with a lasing diameter of 1 mm. First, we have predicted that an intrinsic spectral linewidth of sub kHz and an output power of ~5 W can be simultaneously realized in a PCSEL with a 1-mm lasing diameter by numerical simulations based on a time-dependent three-dimensional coupled-wave analysis considering carrier-photon interactions and a nonuniform temperature distribution. Then, we have experimentally demonstrated a 1-kHz-class intrinsic spectral linewidth, which is determined by the measurement limit of our spectral linewidth measurement system, and an output power of ~5 W in a fabricated PCSEL with a 1-mm lasing diameter under continuous-wave operation. We expect that even narrower spectral linewidths (<10~100 Hz) can be realized by further increasing the lasing diameter of the PCSEL (for example, to ≥ 3 mm) [16,17] as well as by developing a stabilized low-noise current source that can inject current larger than several tens of amperes. Our narrow-linewidth, high-power PCSELs are expected to contribute to the development of various applications, including coherent optical communications and optical sensing such as long-distance free-space optical communications in space and spaceborne light detection and ranging.